\documentclass[prl,twocolumn,showpacs,amsmath,amssymb,superscriptaddress]{revtex4}
\usepackage{graphicx}

\usepackage{ulem} 
\usepackage{dcolumn}
\usepackage{bm}

\usepackage{fancyhdr} 

\begin{document}

\pagestyle{fancy}
\lhead{}
\chead{}
\rhead{}
\lfoot{}
\rfoot{}
\cfoot{\thepage}

\setlength\topmargin{ -0.5in}
\setlength\headsep{ 10pt}
\setlength\footskip{ 30pt}
\setlength\textheight{665pt}
\setlength\textwidth{518pt}
\setlength\parindent{0.13in}
\setlength\parskip{0pt}


\title{The influence of a single quantum dot state on the characteristics of a microdisk laser}

\author{Z. G. Xie} 
\affiliation{Department of Applied Physics, Stanford University, Stanford, CA 94305, USA}
\author{S. G\"otzinger}
\affiliation{Edward L. Ginzton Laboratory, Stanford University, Stanford, CA 94305, USA}
\author{W. Fang}
\affiliation{Department of Physics \& Astronomy, Northwestern University, Evanston, IL 60208, USA}
\author{H. Cao}
\affiliation{Department of Physics \& Astronomy, Northwestern University, Evanston, IL 60208, USA}
\author{G. S. Solomon$^{\ast,}$}
\affiliation{Edward L. Ginzton Laboratory, Stanford University, Stanford, CA 94305, USA}
\affiliation{Solid-State Photonics Laboratory, Stanford University, Stanford, CA 94305, USA}
\affiliation{Atomic Physics Division, NIST, Gaithersburg, MD 20899-8423 USA}
\email{glenn.solomon@nist.gov}


\widetext

\begin{abstract}
We report a quantum dot microcavity laser with a cw sub-$\mu$W lasing threshold, where a significant reduction of the lasing threshold is observed when a single quantum dot (QD) state is aligned with a cavity mode. The quality factor exceeds 15 000 before the system lases. When no QD states are resonant, below threshold the cavity mode initially degrades with increasing pump power, after which saturation occurs and then the cavity mode recovers. We associate the initial cavity mode spoiling with QD state broadening that occurs with increasing pump power.
\end{abstract}


\maketitle
 \narrowtext
While the first masers and lasers were demonstrated in
gas and solid-state systems, respectively, the scaling of
masers \cite{1} and lasers \cite{2} to a small number of emitters
has been investigated for two decades using atomic emitters,
such that recently a one-atom laser has been demonstrated
\cite{3}. These systems feature external cavities where
the atom is typically strongly coupled to the radiation field,
and require ultra-high vacuum chambers. 
{Recently, approaches are 
combining microcavities and atomic emitters \cite{4}; however, they are difficult to implement and are not
monolithic.
 
Semiconductor-based lasers are compact, intrinsically
monolithic, and ubiquitous, but a single-emitter
laser, an analog to the single atom laser, has not been
observed. Typical devices are 
pumped nonresonantly through carrier transport
and employ heterostructures, often with active regions of
reduced dimensionality. An active medium of QDs provides
an ensemble of sharp excitonic states for lasing \cite{5},
and was demonstrated using a dense ensemble of QDs in
1994 \cite{6}. If the QD distribution is made dilute so that only
a few discrete QD states are present, extremely low threshold
lasers can be made provided the photon storage time is
adequate to ensure stimulated emission. Such a system is a
promising route to solid-state, single-emitter lasing.
 
QD microdisk lasers have been reported \cite{7,8}, with
thresholds of 5-20 $\mu$W. In these devices many QDs are
randomly distributed and thus most are at best weakly
coupled to the cavity mode unless additional techniques
are used to align the emitter and cavity mode \cite{9,10,11}. For
single-emitter devices, the number of coupled emitters can be decreased by reducing the cavity size or by using crystal growth techniques that increase
the QD spectral distribution and reduce the overall
density. Single QD
emission can be coupled to micropillars \cite{12}, microdisks \cite{12a}
and 2D photonic crystal cavities \cite{13}. The mode volume in microcavities can be
reduced to a few cubic wavelengths, with cavity {\it Q'}s exceeding
10,000 \cite{14,15,16,17,18}. Here, we demonstrate
a QD microdisk laser with thresholds in the
500 nW range on 1.8 $\mu$m diameter disks with a cavity $Q$
exceeding 15 000 at threshold. While the lasing process likely
involves many emission states, the tuning of a single QD
emission state through the cavity alters the lasing threshold
by a factor of approximately 3.
 
Our samples are grown by molecular beam epitaxy. The
structure contains an initial GaAs layer of 300 nm on a
GaAs substrate, followed by 700 nm of Al$_{0.8}$Ga$_{0.2}$As,
75 nm of GaAs, InAs QDs, and 75 nm of GaAs. The QD
density is approximately 50 QDs/$\mu$m$^{2}$. Microdisks are
fabricated with standard photolithography and an isotropic
wet chemical etch to define an initial post, after which a
HCl-based selective etch is used to undercut the AlGaAs.
For photoluminescence (PL) measurements, the sample
is mounted in a continuous-flow helium cryostat. A 0.75 NA objective
lens is used to collect PL emission from the microdisk
sample. A Ti-sapphire laser operating cw or
mode locked was used to excite the QDs nonresonantly
through the same objective lens. Unless otherwise noted,
measurements were conducted at 7 K with 780 nm cw laser
excitation. Under these conditions, carriers are created in the
GaAs, diffuse and relax through wetting layer states, and
into QD states. The QD emission is fiber-coupled
to either a spectrometer, a
spectrometer-streak camera for wavelength-selective time-resolved
measurements, or a second-order correlation
setup for photon statistics measurements. The beam diameter
of the pump can be adjusted at the sample from 1.5
to 8.5 $\mu$m.
 
PL was first measured from a nonprocessed planar region
of the sample with a large beam size (8.5 $\mu$m) and an
excitation power density of 2.5 W/cm$^{2}$. With this spot
size, there are approximately 3000 QDs illuminated. PL 
from this region is shown in Fig. 1(a). We observe broad,
ensemble QD emission centered around 905 nm. Similar measurements were done on the disk
sample and are shown in Figs. 1(b) and 1(c). However,
because of the small disk areas and hence the small number
of illuminated QDs, the PL spectra show much sparser
overall emission with the presence of discrete emission
states. Second-order correlation measurements show that
these states are typically antibunched. Microcavity whispering-gallery modes
are observed in both figures. The free spectral range of the
cavity is approximately 45 nm in the vicinity of the QDs
emission, and thus two modes are typically observed. One
of the modes (mode L) is located at the long-wavelength
side of the QD distribution, while the other (mode S) is
located at the short-wavelength side. In most of our cavities 
here is a mode splitting of  ~0.2 nm (~300 meV)
[as seen in Figs. 1(b) and 1(c)] of the ideally degenerate counterpropagating modes. The splitting is due to coupling of the modes through 
surface defects or other symmetry breaking in the microdisks.
We identify these as either the long or short wavelength
branch of the mode.
\begin{figure}[b!]
\vspace*{-0.15in}
\begin{center}
		\includegraphics[width=3.2in]{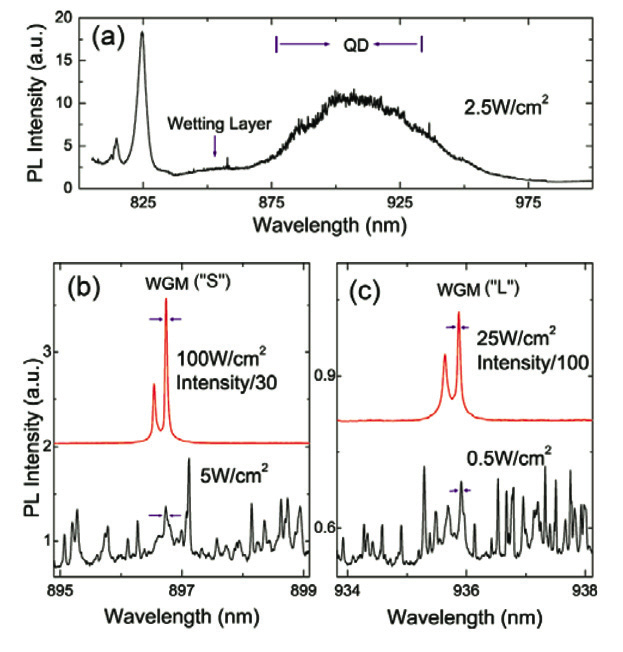}
\caption{(a) PL spectrum from an unprocessed region
at 7 K under low excitation, showing QD emission. (b) PL
spectra on a 1.8 $\mu$m diameter microdisk near cavity mode S, for two different excitation powers. (c) On the same disk, the PL
spectra near mode L, for different excitation powers.}
\label{}
\end{center}
\end{figure} 

\begin{figure}[b!]
\vspace*{-0.15in}
\begin{center}
\includegraphics[width=3.2in]{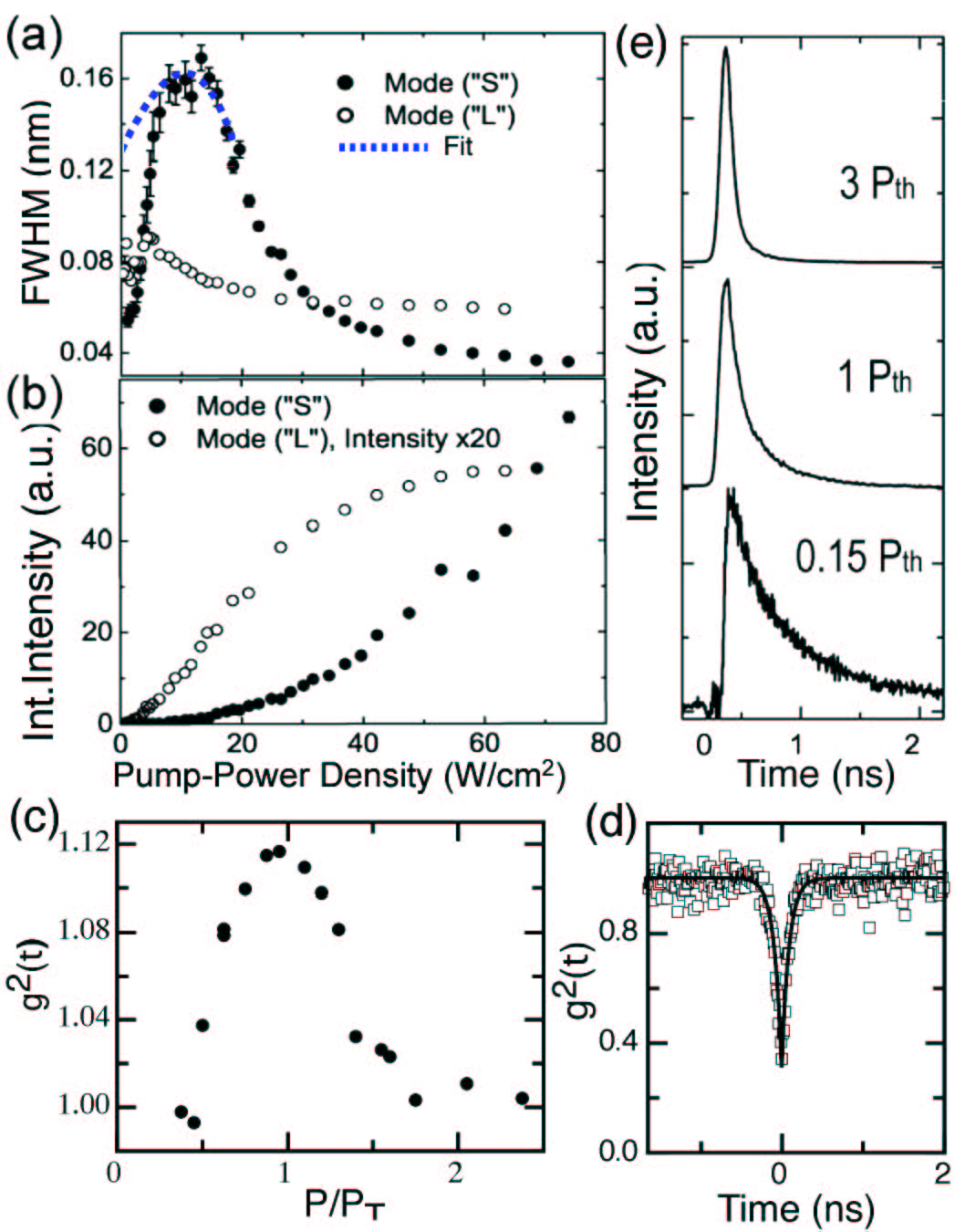}
\caption{(a) Evolution of the linewidth of the
long-wavelength branches of modes S and L shown as a function
of pump power. The lasing threshold is at a power density of
30 W/cm$^{2}$; broken line is theory. (b) The cavity mode integrated
intensity change with pump power shown for S and L;
L saturates before the onset of lasing. (c) Second-order correlation measurements
[g$^{2}(\tau)$] for various pump powers. (d) g$^{2}(\tau)$ in the weak excitation
limit on a QD state separated from the mode indicating
antibunching (no background or dark count correction).
(e) Time-resolved PL measurements show the reduction of the
lifetime associated with stimulated emission from mode S.}
\label{}
\end{center}
\end{figure} 
	We typically observe cavity mode lasing in S but not in
L, likely due to the distribution of states, as seen in
Fig. 1(a). Results of laser pump-power versus intensity (L-I)
measurements on a 1.8 $\mu$m diameter disk are shown in
Figs. 2(a) and 2(b) for the long-wavelength branches of
modes S and L. For mode S, as we increase the excitation
power we first observe a modal linewidth increase from
approximately 0.05 to 0.16 nm (245 $\mu$eV), then a linewidth
decrease to the spectrometer resolution limit of
0.036 nm (55 $\mu$eV). The decrease of the modal
linewidth with increasing pump power is attributed to a
gradual saturation of absorbing states and the onset of
stimulated emission.We performed the same measurement
on the long-wavelength side of mode L. A similar trend is
seen in the linewidth evolution; however, both the initial
modal linewidth broadening and the following decrease are
much smaller, and most significantly, the eventual linewidth
levels off at 0.06 nm, well above the spectrometer
resolution. The L-I curves shown in Fig. 2(b) suggest a
soft lasing transition for mode S, as would be expected for
high spontaneous emission coupling observed in microcavities
\cite{19}. The transition to lasing is confirmed in the
microdisks through second-order correlation measurements,
[Fig. 2(c)] where bunching is observed around
the threshold and Poissonian photon statistics are observed well above
threshold \cite{20,21,22}. For comparison, the L-I curve for
mode L has a nearly linear dependence until the QD states
saturate, indicating the absence of lasing. In comparing the
above data we conclude that the transparency window is
reached at a pump-power density of 30 W/cm$^{2}$ for S, corresponding to
a 750 nW threshold for this microdisk.

The initial broadening of the modal linewidth in
Fig. 2(a) is related to increasing absorption: with increasing
power the QD states broaden, resulting in enhanced
spectral overlap with the cavity mode and enhanced absorption.
At low pump power, the linewidth of the QD states are typically below
the spectrometer resolution. The isolated emission usually shows antibunching in the
second-order correlation function, as is expected from a
single anharmonic quantum system. An example is shown
in Fig. 2(d). However, at threshold the states can broaden to
approximately 0.13 nm (200 $\mu$eV).

To observe the dynamical behavior we have made
wavelength-selective time-resolved measurement on S.
The sample is probed with pulsed 850 nm laser light, in
the vicinity of the InAs wetting layer to reduce the carrier
relaxation time to the QD states. Under this condition the absorption
coefficient is roughly 20 times lower. A selection of our pump power
results is shown in Fig. 2(e), where the emission is
on resonance with cavity mode S. At a low pump power,
the time-resolved intensity curve shows the emission decay
of excitons with a lifetime of $\sim$500 ps. With
increasing pump power, the measured lifetime decreases,
and finally reaches a decay time of 47 ps. Since we observe
a rise time of $\sim$50 ps due to carrier diffusion
to and relaxation into the QD states, the 47 ps decay time is
now determined by a combination of the carrier capture
processes, the lifetime, and our $\sim$20 ps temporal
resolution.

From statistics, we expect 130 QDs inside
the disk plane and approximately 60 QDs spatially
located in the cavity mode region, which extends $\sim$
250 nm into the disk plane from the edge. On
average, the number of QD states spatially and spectrally
coupled to the cavity resonance at low excitation power is much less
than unity. However, the actual spectral coupling near the
lasing threshold is determined by the homogenous broadened
linewidth of the QD emission.

\begin{figure}[b!]
\vspace*{-1.2em}
\begin{center}
		\includegraphics[width=3.2in]{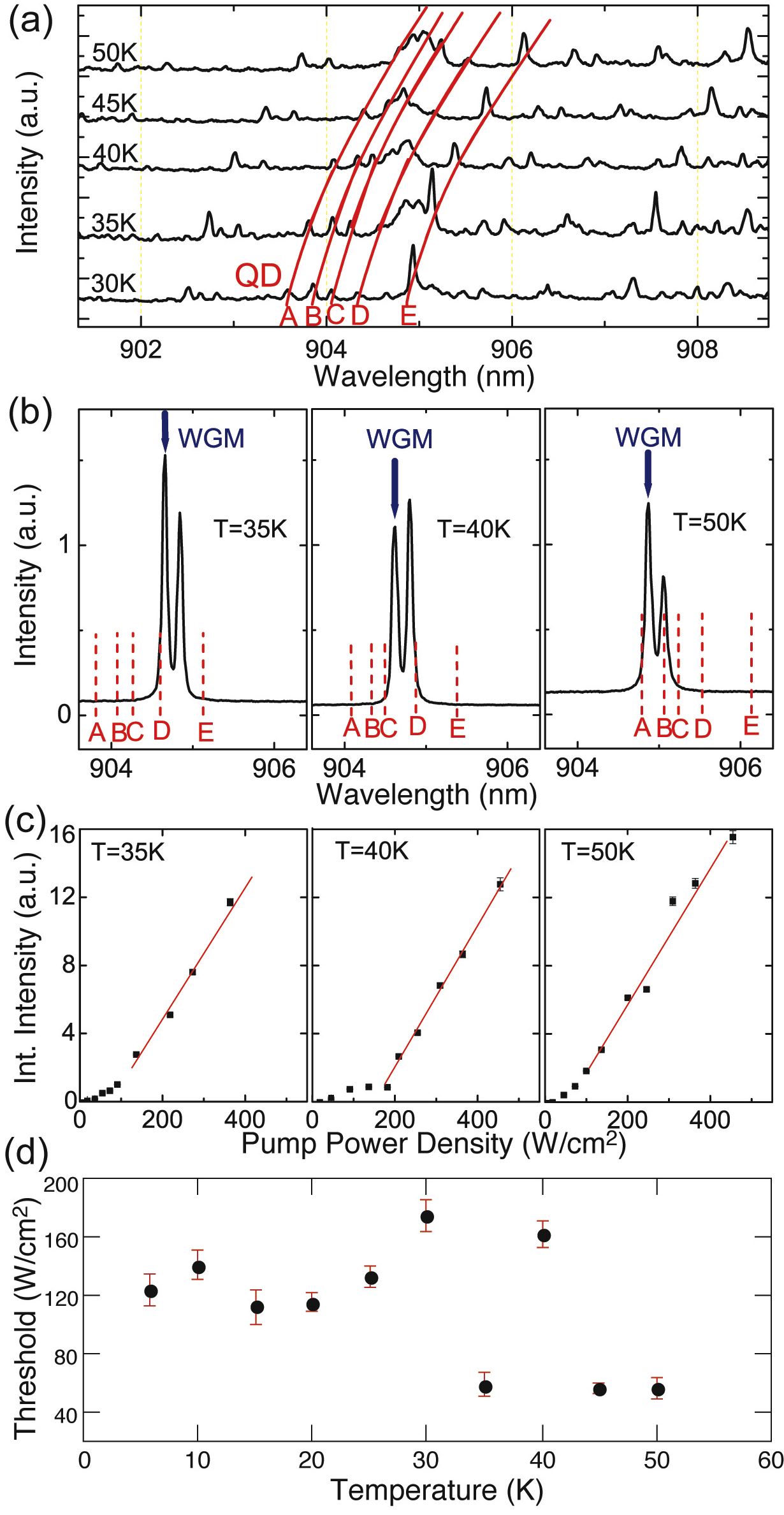}
\caption{1 (a) PL spectra of QD emission and mode
position at different temperatures. Between 35 to 50 K, QD lines
A, B, C, and D cross the cavity modes. Lines are guides
following QD emission shifts. (b) The PL spectra show the
spectral alignment of QD A, B, C, and D (under weak
excitation) with mode S at high excitation. At 35 K D is aligned and
at 50 K A is aligned with the short-wavelength branch of the split mode.
(c) The L-I curves at 35, 40, and 50 K. Lines are guides.
(d) Lasing thresholds at various sample temperatures.}
\label{}
\end{center}
\vspace*{-0.9em}
\end{figure} 

To clarify the modal linewidth broadening and lasing
with the apparent absence of an aligned QD state we suppose
there are several QD states in the spectral vicinity of the
cavity mode, each with average occupancies $N_{i}$, where $0\leq N_{i} \leq1$ for each QD state $i$. If we assume that states from different QDs do not
interact, a general description of the net stimulated emission
or absorption is $\sum _{i} \beta_{i} \dot g_{i} \left( N_{i} - \frac{1}{2} \right)$, where $g_{i}$ is the
coupling factor between the $i$th QD state and cavity
mode, and $g_{i}$ is the stimulated emission (absorption) coefficient for the $i$th state. Because the spatial and polarization
alignments of the cavity mode and each QD state $i$ are constant, $\beta_{i}$ is
equivalent to the spectral coupling factor modified by a
constant, and is $\beta_{i} = \alpha_{i} \int {f \left( \lambda - \lambda_{0}, \Delta \lambda_{0}\right) h \left(\lambda - \lambda_{i}, \Delta \lambda_{i}\right)} \rm{d} \lambda$, where $f \left( \lambda - \lambda_{0}, \Delta \lambda_{0}\right)$ is the spectral distribution
of the cavity mode, $h \left(\lambda - \lambda_{i}, \Delta \lambda_{i}\right) d\lambda $ is the spectral
distribution of the $i$th state, and $\alpha_{i}$ is a constant.  
 $\lambda_{0}$ and
 $\lambda_{i}$ are the wavelengths of the cavity mode and $i$th state, and
 $\Delta \lambda_{0}$ and $\Delta \lambda_{i}$ are their respective linewidths. 
 Assuming
Lorentzian distributions for $f$ and $h$,  $\beta_{i} \propto \pi^{-1} \left( \Delta \lambda_{0} + \Delta \lambda_{i} \right) \times [ \left(4 \lambda_{0} - \lambda_{i} \right)^{2} +  \left(\Delta \lambda_{0} + \Delta \lambda_{i} \right)^{2} ]^{-1}$.
Since we observe that  $\Delta \lambda_{i}$ is pump-power dependent, $\beta_{i}$ will be as well. 
Below transparency,
$\Delta \lambda_{0} \propto \sum _{i} g_{i} \beta_{i} ( N - \frac{1}{2} )$. To estimate the change in $\Delta \lambda_{0}$ with pump power, we assume only the emitters nearest 
the cavity mode couple, and use the average nearest neighbor emitter modal detuning, with a detuning taken 
from experimental data ($\sim$0.25 nm). We use a value for $\Delta \lambda_{i}$ at threshold taken from the data (0.13 nm) and assume a sublinear excitation dependence \cite{23}. Finally, we assume
$N_{i}$ varies linearly with pump power below transparency. Then, the modal linewidth variation can be determined with only a scaling 
factor as a free parameter. This is shown as a broken line in
Fig. 2(a), and accurately represents the linewidth broadening.
Thus, when QD states are present and subject to
power broadening, but are initially only weakly overlapping with
the cavity, we can expect themodal linewidth to
increase below transparency. Furthermore, the system can lase at
higher powers if the cavity $Q$ is adequate. Now, we contrast
this to the situation when a QD state is aligned with the
cavity mode.

Because the refractive index change with temperature is
much smaller than the bandgap change with temperature,
the QD emission lines can be tuned through the cavity
modes by adjusting the sample temperature. In most of our
samples lasing persists when the sample is tuned from 6 to
55 K (a QD tuning range of 1.5 nm). This indicates the
lasing is not based exclusively on observable QD states
resonantly coupled to the mode \cite{22}. However, the relative
spectral tuning of observed QDs emission states and cavity
modes does influence the L-I curve. In Fig. 3(a), we show
the position of cavity modes and QD emission peaks when
the sample is temperature tuned under low pump power. At
35 K the QD emission line D crosses the cavity mode, and
at 50 K the QD emission lines A and B also cross the mode.
Because the QDs are randomly distributed throughout the
disk, spectral alignment between the cavity mode and QD
state does not guarantee coupling. We show spectra and
L-I measurements for the short-wavelength branch of
mode S at three different sample temperatures, 35, 40,
and 50 K in Figs. 3(b) and 3(c). In Fig. 3(b), we show
the PL intensities of the split mode S at a pump 
level where both branches are lasing. Because of
the large emission from the cavity modes, individual QD
emission lines can no longer be identified; however, we
mark the emission states present under weak excitation
[see Fig. 3(a)]. In Fig. 3(c), the lasing threshold of the
left branch of mode S is approximately 3 times lower at
50 K than its threshold at 40 K. At 50K, when the pump power is
increased so as to pass through the threshold of the left
branch, the ratio of integrated intensities
of the left-to-right branches changes from 0.2 to 1.6. We
explained this by the selective coupling of the QD emission
state A with the left branch of the mode. Similarly, at 35 K,
emission line D is closely aligned with the short wavelength
branch of mode S, and the threshold is also
significantly reduced with respect to 40 K. The threshold
values as a function of temperature are summarized in
Fig. 3(d). We observe an average of approximately 3 times
reduction in threshold when a QD state is aligned with the
cavity mode.

We have shown a solid-state laser in which a single
emitter, here a single QD state, plays a significant
role.  Effects due to isolated emitters are seen below and at threshold. Below threshold the initial
broadening of the cavity mode is attributed to QD state
broadening. When a single QD emission line is tuned to the
resonance, the threshold is decreased by a factor of approximately
3. However, under many conditions the cavity
lases even though no emitter emission is observed on
resonance, as was recently observed in Ref. \cite{22}. To
achieve single state lasing the processes associated with
the loss must be suppressed and more efficient lasing via
the single-emitter state (i.e., higher oscillator
strength and higher Q), must be implemented.

We thank Y. Yamamoto for fruitful discussions and use
of equipment, and B. Zhang for MBE assistance. Z. G. X.
thanks J. S. Harris for advice and encouragement. S. G.
acknowledges the Alexander von Humboldt Foundation
for financial support.

\end{document}